\documentstyle[prl,aps,floats,epsf,epsfig]{revtex}

\begin{document} 
\draft

\title{A Possible Via For Proving The Approach Of The Hubbard Ground
States
Toward Stripes And Other Doping States}

\author{Zohar Nussinov}
\address{Institute Lorentz for Theoretical Physics, Leiden
University\\ P.O.B. 9506, 2300 RA Leiden, The Netherlands}

\twocolumn[

\widetext
\begin{@twocolumnfalse}

\maketitle

\begin{abstract}

We suggest a simple way for rigorously establishing constraints
on the form of the ground states of the Hubbard and t-J models
and extended longer (yet finite) range variants for various 
dopings, once ``exact'' numerical results
are established for these Hamiltonians on finite size 
clusters with two different
(both open and periodic) boundary conditions.
We demonstrate that strong bounds will be established
if non-uniform minima subject periodic boundary
conditions are found. 
An offshoot of our proposal might enable
rigorously establishing the strong tendency toward especially 
stable stripe formation at the magical commensurate doping of 1/8 on the
infinite lattice, as well as at other dopings.

\end{abstract}

\vspace{0.5cm}

\narrowtext

\end{@twocolumnfalse}
]

\section{Introduction and Outline}

The contents of this very simple note 
amount to the observation that 
finite size numerical 
calculations can yield 
precise information on
the form of the ground state 
on the infinite lattice.
In particular, we will
ask what might be stated 
if two exact numerical computations
find non-uniform minima when
both subjected to both open 
and closed boundary 
conditions.  We will show
that in such instances,
the majority of the 
finite size fragments
of the true ground
state extending over the infinite 
lattice will display inhomogeneities
intermediate between those 
obtained in the open and 
closed boundary condition
miniminization on the 
finite fragment. Though
the considerations that
we have in mind are very
general and apply to
any finite range model, we will 
specifically address 
the two dimensional 
Hubbard and related models. 
The well known Hubbard 
model is given by

\begin{equation}
\label{Hubbard model}
H=-t\sum _{\langle ij\rangle, \sigma }(c_{i\sigma }^{\dag }c_{j\sigma
}+c_{i\sigma }^{\dag }c_{j\sigma })+U\sum _{i} n_{i\uparrow }n_{i\downarrow},
\end{equation}%
where \( c_{i\sigma }^{\dag } \) creates an electron on site \( i \)
with spin \( \sigma  \) and \( j \) is a nearest neighbor of \( i \).
This model contains both the movement of the electrons (hopping) (\( t \),
kinetic energy) and the interactions of the electrons if they are
on the same site (\( U \), potential energy). 
The Hubbard model is one of the simplest possible models 
of interacting electrons. In the large $U$ limit,
the model reduces to the well known t-J model
\begin{eqnarray}
H =
- t \sum_{\langle i j \rangle, \sigma}  (c_{i, \sigma}^{\dagger} c_{j, \sigma}
+ H.c.) +  J \sum_{\langle i j \rangle}  \vec{S}_{i} \cdot \vec{S}_{j}
\end{eqnarray}
where $\vec{S}_{i} = \sum_{\sigma \sigma^{\prime}} c_{i
\sigma}^{\dagger} \vec{\sigma}_{\sigma, \sigma^{\prime}} c_{i,
\sigma^{\prime}}$
is the spin of the electron at site $i$,  
$\vec{\sigma}$ are the Pauli matrices, and there is a constraint
of no double occupancy of any site $i$ ($n_{i} = \sum_{\sigma}
c_{i,\sigma}^{\dagger}c_{i,\sigma}$ has expectation values  0 or 1).

Numerical calculations on the t-J model
with open and cylindrical boundary conditions
have found non-uniform ``stripe'' 
patterns \cite{WS} which have earlier been predicted
earlier by approximate 
solutions to the Hubbard model \cite{stripes},
and argued for, very convincingly,
by the addition of the strong Coulomb
effects present in the cuprates \cite{phase}. 
At the moment, there 
is no clear consensus as to whether these patterns display the
genuine ground state of the bare Hubbard (or t-J) Hamiltonian
or amount to a finite size artifact. We will argue
that if periodic boundary condition calculations
reproduce a similar periodic pattern for the
bare Hubbard model or its more physical extension
containing a few additional finite range 
Coulomb contributions then the existence
of stripes is essentially proved for 
the infinite lattice within bounds 
that we will derive.

The outline is as follows: In section(\ref{sec-rep}),
we will introduce a simple discrete representation
for depicting the charge and spin densities at various
sites or plaquettes. All of the bounds that we derive will
pertain to the way in which the global ground
state tiling the infinite lattice will look 
in this representation. Our motivation for
employing a discrete representation
is obtain energy spectra for each
representation whose minima are
separated from each other by 
finite gaps. Next, in section(\ref{sec-gb}),
we label the various possible discrete
configurations on a finite size fragment as 
``good'' or ``bad'' depending on whether 
they appear as the ground state minima
in various cases. If the numerical calculation
on the finite size fragment reflects the 
true state of affairs, then in covering
the true ground state on the infinite 
lattice we will find that within
most finite size fragments the discrete
configurations look much the same (up to trivial
translations) as those suggested by the finite
size numerical calculation. Ideally, 
each finite size fragment will look 
like that suggested by the numerical
calculation (and will be ``good''). In this short
note we derive a bound on the number
of ``bad'' textures that appear in the 
true ground state on the infinite lattice
but are not anticipated from the 
finite size calculations (i.e. 
are not finite size minima).
The actual bound that we derive 
is explicitly stated in section(\ref{what}).

In section(\ref{means}), we quickly outline the
path that we will follow in order to 
derive these bounds. In section(\ref{lower-sec}),
we derive lower bounds on the
ground state energies in the various
discrete sectors by examining the open
boundary condition problem. In section(\ref{upper-sec}),
trivial upper bounds on the global ground state energy density
are attained by looking at the periodic boundary condition
problem. In section(\ref{pieces}), we fuse our
lower and upper bounds to derive a bound on 
the number of ``bad'' blocks appearing in
a covering of the global ground state minima
which are not anticipated from finite
size calculations. It is important 
to emphasize that as the discrete representation
of the ground state is up to the user,
we do not need to examine matters
in detail. As discussed in section(\ref{why}),
any single given calculation on 
the finite size system with 
open and closed boundary condition
is sufficient to prove 
that in most finite size blocks
the charge and spin expectation
values will interpolate, in one 
global ground state spanning
the entire lattice, between the 
values obtained in both
numerical problems. Although
the particular problem that we have
in mind is that of stripes, these
simple considerations are very general.

\section{A discrete representation}
\label{sec-rep}

\begin{figure}
\includegraphics[width=5cm]{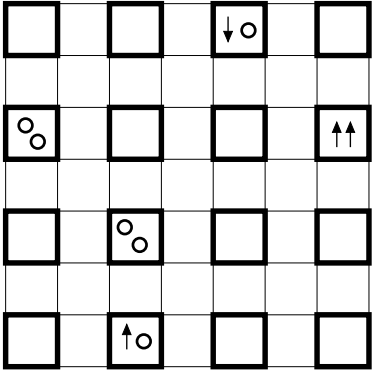}
\caption{This figure is reproduced
from Altman and Auerbach. Each plaquette on
one dual sublattice is in a certain 
coarse grained state. The two and one 
hole states along with spin triplet and
singlet states are clearly seen. The very interesting
work of Altman and Auerbach focuses 
on the plaquettes (non overlapping minimal clusters) 
residing on only one sublattice- thereby breaking 
translational invariance.
In the current note, we 
consider covering the lattice
with all {\em maximally overlapping}
clusters $\{\lambda\}$, wherein 
translational symmetry
is unbroken.}
\label{AAA1}
\end{figure}

We can coarse grain a two dimensional
Hubbard system
to ask the state of each individual plaquette. This is schematically
illustrated for half of the plaquettes in Fig.(\ref{AAA1}), 
which is reproduced from \cite{AAA}. 
In what follows
we will endow the electronic states with discrete bar code
patterns specifying the coarse grained electronic states
within each plaquette. 

The bar code specification, detailed below, is 
completely up to the user. Given a wave-function
$|\psi \rangle$ on the infinite system or finite
size cluster, we compute the number and
total spin expectation values
for each individual plaquette on which
the wave-function has support. To be more 
precise, if $R$ is a plaquette in the finite
cluster $\lambda$ or in the infinite lattice $\Lambda$,
with vertices labeled by $R_{1 \le i \le 4}$, 
then the net charge and squared spin of $R$ are 
\begin{eqnarray}
\langle \psi| \sum_{i}  n_{R_{i}}  | \psi \rangle  \equiv n_{R} 
\end{eqnarray}
\begin{eqnarray}
\langle \psi | [\sum_{i} \vec{S}_{R_{i}}]^{2} | \psi \rangle \equiv S_{R}(S_{R}+1).
\end{eqnarray}

We might specify plaquettes with occupancy $n_{R}$, to have
an integer number of $[n_{R}+1/2]$ holes, with $[...]$ the integer
part ``function''. If greater accuracy is desired
we may quantize the value of bar-coded charge in multiples
of half a hole etc.  We may similarly, arbitrarily
decide to call the projected spin states within the 
plaquette as singlet, triplet etc.
according to the appropriately rounded off 
value of $S_{R}$. Of course, the user may decide
to round off the number density according to
his or her own whim and based on the value of $n_{R}$, 
decide that $R$ is  
a plaquette with no holes, one holes, 
two holes etc.  The bar code representation,
does not do justice
to the rich Hubbard states. This mapping,
in the RG spirit, is not invertible.
All of our proposed bounds pertain 
to bounding the frequency of these 
various bar code patterns.

Now, suppose, that we want to see if
stripes are formed, or if some other
interesting states might emerge (or not)
for various dopings. In the
coarse grained representation, 
this amounts to seeing charge
and spin nestled in the corresponding 
plaquettes compromising
these configurations.
The cutoff criteria
for rounding off the number
and spin expectation values $n_{R}$ and $S_{R}$ 
might be readjusted to see most crisply the 
patterns that we wish to probe. 

As the astute reader might have guessed,
the purpose of focusing on abbreviated discrete
coarse grained representations of the 
Hubbard states is that, the bottom of the energy spectra 
of discrete 
bar code sectors tend to be separated by 
finite gaps. Finite gaps are usually a good
thing when we wish to prove the 
occurrence or non-occurrence of certain
configurations.

If we
wish to probe only the charge degrees of
freedom, we may examine the occupancy 
of each individual site (and no longer 
focus on plaquettes) to see if it is occupied
by, say, half a hole, or by a hole 
or by nothing whatsoever. 

The bar code specification of the states need not
be uniform. For instance, in what follows, to better incorporate
open boundary condition aberrations of the ground states 
close to the boundary, we may partition the finite cluster 
$\lambda$ into plaquettes everywhere deep inside the 
cluster and take a coarser covering (e.g., by a three by three  
plaquette cluster) toward the edges:
the bar code of the coarser covering toward the edges
will be more loose and will enable more states. The 
coarse covering and rounding at the edges 
can be arbitrarily loose (it may uniformly set to
a bar code symbol enabling all possible
spin and charge distributions within) to 
effectively excise a finite thickness halo surrounding 
the boundary of $\lambda$. 
The reader can invent 
and improve such 
definitions.

\begin{figure}
\includegraphics[width=9cm]{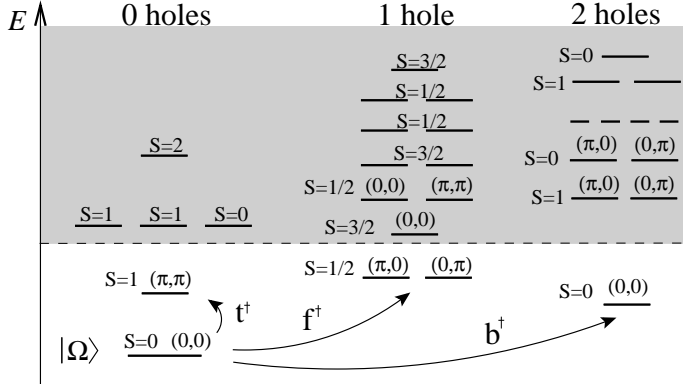}
\caption{This figure is reproduced from 
Altman and Auerbach. The diagram
displays the exact results from the exact diagonalization
of a single plaquette. Aside from the undoped plaquette
with singlet correlations (the vacuum $|\Omega \rangle$), 
the three lowest lying 
excited states (the triplet, the low energy fermion (the single
hole), and the low lying boson (two holes)) are 
the the states of most
relevance in the low energy sector 
of problem. In the low energy
world, having a certain bar code 
pattern generally goes hand in hand with a
high amplitude for the corresponding low
energy eigenstates of the pattern associated
with each of the individual plaquettes.}
\label{ladder1}
\end{figure}

\section{A simple classification}
\label{sec-gb}

Let us imagine that numerical results
hint very strongly at a certain 
abbreviated ground state of the Hubbard or t-J
model (e.g a bar code representation of the DMRG results
of White and Scalapino along with a coarser covering
toward the edges to better remove the effect of 
boundary conditions toward the edges). If a certain 
periodic abbreviated ground state appears
with some period $L$ then the appearance of
such a state and its cousins generated by translations (relying 
on the observed
periodicity) on a finite $L \times L$ (or larger) 
slab $\lambda$ will be termed ``good''. All other states
not related by symmetry and not appearing in
the numerical ground state will be 
defined as ``bad''.

If the numerical results indeed reflect the
true state of affairs and are not an artifact
of spurious boundary conditions or other finite size 
effects, then in tiling the real physical infinite 
lattice $\Lambda$ with all maximally overlapping 
windows $\{\lambda\}$, we should observe many more ``good''
states than ``bad''.

\section{What are we after?} 
\label{what}

In a nutshell, we wish to show how we 
can bound the number of ``bad'' bar coded
states in all maximally overlapping
clusters $\{ \lambda \}$ vis a vis the
number of the ``good'' candidate states.

The bound that we will momentarily
arrive at trivially reads
\begin{eqnarray}
N_{b} \le (\frac{\delta}{\Delta}) N_{g}. 
\end{eqnarray}

The quantities $\delta$ and $\Delta$ are gaps 
that we will detail below.

\section{The Means}
\label{means}

Although not the most efficient, we will
examine the energy of the system. The two
gaps (or gap and anti-gap) $\Delta$ and $\delta$
are tied to the energy of the cluster $\lambda$ when 
subjected to open and periodic boundary conditions
respectively. $\Delta$ denotes by (at least) how 
much the energy of any bar code configuration
apart from the bar code sector of the true
numerical ground state on the finite size
cluster is elevated relative to the
absolute energy minimum on the finite
size cluster $\lambda$. 

As will be shown below, a trivial
lower bound on the energy of the system
may be obtained by covering the entire
lattice $\Lambda$ with all maximally
overlapping finite blocks $\{ \lambda\}$,
consequently registering the bar code representation
within each block $\lambda$, and finally summing
the lower bounds on the energies of each of the 
bar code patterns that appear while tiling the system
with all these maximally overlapping
windows $\{ \lambda \}$. The gap $\Delta$ 
computed by examining constrained open boundary
condition minima
is the lower bound on the energy increase
each time a ``bad'' bar coded pattern is 
encountered.

A trivial upper bound on the true ground state energy
of the infinite system may be obtained by merely
minimizing the energy on a $\lambda$ subject to 
periodic boundary conditions. Obviously, the
open boundary condition minimum will be lower (or the same)
when compared to its periodic boundary condition counterpart.
The difference between the two is denoted 
by $\delta$. As the size of the cluster $\lambda$
becomes larger and larger, the periodic boundary
condition and open boundary condition must
veer toward each other. Small delta ($\delta$) is 
indeed small for large enough $L$.

There is one detail which we have so far 
tucked away under the rug- in evaluating 
the energies on finite size cluster $\lambda$
we must renormalize the parameters $t, U, ...$ such
that the sum of the Hamiltonians on each individual cluster
\begin{eqnarray}
H = \sum_{\{ \lambda \} } h_{\lambda} 
\end{eqnarray}
is the global Hamiltonian on the entire system $\Lambda$. 
The number of nearest neighbor pairs differs from the
number of the number of sites and consequently the parameters
associated with these (t and U respectively)  
must be readjusted. A similar occurrence happens if longer
range (yet obviously finite) interactions are introduced.

\section{Translational Symmetry}
\label{trans}

Translational symmetry is not
broken in the scheme presented
here. The reader will note
that we examine all 
maximally {\em overlapping} blocks
$\{\lambda\}$ in concert. We are
not looking at a specific 
subset of clusters- that 
would indeed break translational symmetry.

This maximally overlapping
covering is a bit non-traditional.
Virtual hole evaporation and other
processes that one might worry 
will not taken into account
when examining small fragments
of the cluster are very transparently 
accounted for by this and many other coverings. The
maximal overlapping covering
is more important for easily 
incorporating correctly longer range
interactions (next nearest neighbor and beyond) 
The reader can invent a multitude
of other variants.

\section{Chemical Potential and
Doping}
\label{chem}

Throughout the minimization
procedures, the number of holes 
is {\em not} held fixed over each
of the small clusters $\lambda$. 
A chemical 
potential term is inserted
instead to provide the correct
density of holes over the entire 
lattice $\Lambda$. The renormalization 
of the chemical potential term
with the size of the block $L$ 
is identical to the scaling of 
the on-site Hubbard
repulsion term $U$.

\section{Lower Bounds (via Constrained Open Boundary Condition Minima)}
\label{lower-sec}

Here we give the reader a flavor 
of how constrained lower bounds may 
be easily generated and employed.
In order to set the stage for things
to come and to distill
the essentials, 
we will start
by examining the situation
when the cluster $\lambda$ is the 
minimal single 
plaquette. Later on we will
show how all of this may be 
trivially extended for larger
covering clusters. 

The state of the system is a linear combination 
of the states in the basis vectors in the 
Hilbert space spanned by 
\begin{eqnarray}
|e_{\alpha}^{\prime} \rangle = | 1^{\prime} \rangle \otimes | 2^{\prime} \rangle ... \otimes 
|N^{\prime} \rangle,
\label{Hilbert}
\end{eqnarray}
where $| i^{\prime} \rangle$
specifies the state of the electron (including. its absence- i.e. a hole) 
at site $i$.

Now, let us imagine duplicating
the basis vectors at every site.
At each site $i$ we assign 
two basis vectors $| i^{A} \rangle$
and $|i^{B} \rangle$.  

Let us furthermore define
the even sublattice basis state by 
\begin{eqnarray}
|e_{A} \rangle = |Plaquette_{0} \rangle \otimes |Plaquette_{2} \rangle \otimes
... 
\end{eqnarray}
where each of the plaquettes $\{ |Plaquette_{2n} \rangle\}$ lies 
in the even (A) sublattice.

Each plaquette basis state vector in the even
sublattice is given by a direct product 
of the electronic basis state vectors at each one of its 
four sites
\begin{eqnarray}
|Plaquette_{2k} \rangle = \prod_{i^{A} \in Plaquette_{2k}} 
|i^{A} \rangle
\end{eqnarray}
where, as before, $|i^{A} \rangle$ denotes 
the individual electronic state at site $i$.

Similar definitions and expressions may be 
written for the odd (B) sublattice.

As we must constrain
\begin{eqnarray}
| i^{A} \rangle = |i^{B} \rangle
\end{eqnarray}
for each lattice site $i$,
the Hilbert space spanned by
\begin{eqnarray}
|e_{\beta} \rangle =  |e_{A} \rangle \otimes |e_{B} \rangle
\label{bi}
\end{eqnarray}
contains the physical Hilbert space
of Eqn.(\ref{Hilbert}) as a 
special subspace. 

Now, let us write the full 
(range one Hubbard or t-J) 
Hamiltonian as 
\begin{eqnarray}
H  = \frac{1}{2}(H_{A} + H_{B})
\label{AB}
\end{eqnarray}
with $H_{A}$ having non-vanishing
matrix elements only within
the $\{|e_{A} \rangle\}$ (which is the same as the 
$\{| i_{A} \rangle \}$) subspace. 
Similarly, $H_{B}$ has matrix elements
only within the $B$ subspace. The factor
of one half in Eqn.(\ref{AB}) was inserted
due to the doubling of degrees of 
freedom. Each electronic state makes an appearance
twice- once in the $\{|i_{A} \rangle\}$ basis and
once in the $\{|i_{B} \rangle\}$ basis. 
Now, note that the decomposition of Eqn.(\ref{AB})
is such that each Hamiltonian on the right hand
side does not connect sites on different 
plaquettes- there are no inter-plaquette
interactions in either $H_{A}$ or $H_{B}$. 
We may write 
\begin{eqnarray}
H_{A} = \sum_{k} h_{2k}  \nonumber
\\ H_{B} = \sum_{k} h_{2k+1}
\end{eqnarray} 
where $\{h_{n}\}$ are Hamiltonians that
reside only in the local basis
of the $n-$th plaquette.

As we increased the size of the
Hilbert space, the minimum 
over the real physical Hilbert
space is bounded from below
by the minimum attained
by its extended replicated 
counterpart:
\begin{eqnarray}
\min_{|e_{\alpha}^{\prime} \rangle} \{H\} \ge \frac{1}{2} \Big[
\min_{|e_{A} \rangle} \{H_{A} \} + \min_{|e_{A} \rangle} \{H_{B} \}
\Big] \nonumber
\\ = \frac{N}{4} [\min\{h_{2k}\} + \min\{h_{2k+1}\}].
\end{eqnarray}

This is a trivial bound on the 
ground state energy of the system.

Suppose now that we were not interested
in the global ground state but rather
a restricted minimum. We can minimize the 
energy of the system  
subject to the constraint
that within each odd plaquette
we have one boson (hole pair) and 
within each even plaquette
we have one hole. As the 
replicated Hilbert space
is larger than the 
physical one (it includes
the latter as a special
subset) we will
still trivially have 
\bigskip

$
\min_{|e_{\alpha}^{\prime} \rangle} \{H\}$ such
that within  each odd plaquette we have
a hole pair and within each even plaquette
we have a single hole $\ge \frac{N}{4}  [\min\{h_{2k}\}$ 
such that we have one hole in the $2k$-th plaquette 
+ $\min\{h_{2k+1}\}$ such that we have a hole
pair in the $(2k+1)$-th plaquette $].$

\bigskip

All that we are doing above is tiling the lattice
with dominos and for each of the two sides 
of the domino ($A$ and $B$) we find 
the minimum.

If we want to prove lower bounds on a more
complicated commensurate structure
(not a checkerboard one as
we just considered now
e.g. one with hole pairs on the red squares and 
single holes on the black squares)
then we can consider clusters instead
of single plaquettes and dominos. 

Suppose that we want to prove that 
the ground state for a certain filling
generates a certain commensurate
pattern which occupies finite $(L \times L)$
blocks (clusters). To do that, we might naively
consider all possible bar code sectors.
However, that is not necessary. 
As we will detail in section(\ref{why}),
it is sufficient to find the 
periodic boundary condition minimum
(whatever it may be) and to 
combine it with the unrestrained
open boundary condition
minimum in order to obtain 
powerful bounds on a properly
defined ``good'' sector that
includes both minima.

The number of naive bar code sectors 
is exponential in the area
of the cluster. This number is lowered by
symmetry. The bulk of these
bar code configurations are likely of
high energy and 
the number
of contenders for the low
lying  configurations
might not be as forbidding.
If we are examining natural
potential candidates for 
the states appearing at 1/8 
doping, the 
commensurability is very low
and the number of various sectors
is very small.
Instead, for merely establishing 
a minimal gap $\Delta$, we can simply 
compute the global minima anywhere 
outside the ``good'' bar code 
sectors and compare to the unconstrained
global minima (occurring,
by definition in the ``good'' 
bar code states.

To proceed with the analysis
of the $(L \times L)$ case, 
we generate $L^{2}$ replicas
of the electronic state $|i^{\prime} \rangle$
at each site. We envision covering the lattice
with all maximally overlapping $(L \times L)$ 
blocks $\lambda$. 
We may next define 
Hamiltonians $\{h_{\lambda}\}$
residing in the local 
subspace of the $L \times L$ block. 
For a fragment (cluster)
of size $L \times L$ plaquettes,
there is a multiplicative
factor of $1/(L(L+1))$ that comes
in the definition of $h_{\lambda}$ 
for a pure nearest neighbor
(e.g. t-J) model when evaluating
the energy of the
entire system
by summing over
the energies in
all maximally
overlapping
$L \times L$ blocks $\lambda$.

Similarly, for the Hubbard
model, in tiling
the entire plane by
maximally overlapping 
small $L \times L$ fragments  $\lambda$
each site (U) is over-counted
$(L+1)^{2}$ times while
each kinetic term (t)
is over-counted
$L(L+1)$ times.
Consequently,
for the Hubbard model,
one will need to
examine the readjusted
values of $t, U$ on
the single $L \times L$ fragment
of plaquettes so
as to produce the
correct energies on
the entire lattice.

By covering the lattice
with all $L \times L$ fragments $\{\lambda\}$
and computing the energy within
each fragment, and summing all 
of the lower bounds on the
energies together, we 
may derive very generous 
lower bounds on the energies 
of the global system when the latter is 
subject to the demand that an $L \times L$ cluster 
if it belongs to a certain bar code sector.

It is important to emphasize that
the bar code notation does not
provide quantum numbers. The bar code
values simply code in for rounded 
off values of the various expectation
values. Each bar code configuration 
corresponds to many possible quantum
states. Each individual state in the real physical 
Hilbert space also makes an appearance in the 
extended replicated Hilbert space (but not 
the converse). As a consequence, in performing a 
constrained minimum over a certain bar code sector
or its complement, any state in the real physical
Hilbert space that satisfies the correct
expectation values bounds on the number occupancies 
etc. also makes an appearance in the replicated
Hilbert space and therefore the 
minimum in the replicated space is
lower (or the same) as that of the 
real unreplicated Hilbert space.

\begin{figure}
\includegraphics[width=4cm]{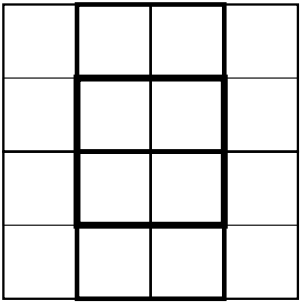}
\caption{A covering of a fragment of the 
 lattice with all maximally overlapping
clusters $\{\lambda\}$. Here the various 
clusters are labeled by varying 
thickness. Each plaquette makes an appearance 
in $L^{2}$ different clusters $\lambda$. In 
the figure above $L =2$.}
\label{ladder1}
\end{figure}

\section{Upper bounds - via The Periodic Boundary Condition Problem}
\label{upper-sec}
 
It is very easy to 
find an upper bound 
on the energy of 
a minimizing $L \times L$ 
configuration when the latter
is embedded in the infinite lattice.
We simply note that
by the variational principle,
\begin{eqnarray}
\langle \psi_{var}| H | \psi_{var} \rangle \ge
\min_{|e_{\alpha}^{\prime} \rangle} \{H\} .
\end{eqnarray} 

Now let us choose the variational
wave-function $| \psi_{var} \rangle$
to be the one 
that minimizes the Hamiltonian 
on an $L \times L$ system 
when subjected to periodic boundary
conditions. Non-uniform 
states can easily (and do in many problems) 
appear as periodic boundary condition
minima; periodic boundary conditions
merely constrain the components of the
allowed wave-numbers to be integer multiples
of $(2 \pi/L)$.
As $| \psi_{var} \rangle$
satisfies periodic boundary conditions,
the $L \times L$ wave-function may trivially cover 
the entire macroscopic lattice.
This state will belong
to some sector. If it does
not belong to the true
macroscopic ground state
pattern, we might need
to increase $L$ until
the periodic boundary condition
minimum will lie in the ``good''
ground state sector of the
system.

Just as in the discussion in the previous
section, we will need to renormalize the
interaction parameters in an open finite block 
$h_{\lambda}$
so that the sum of all these,
\begin{eqnarray}
\sum_{\lambda} h_{\lambda} = H.
\end{eqnarray}
The normalization will be
the same as in 
the previous case of
the derivation of lower bounds.
After all, when tiling the
plane with all possible $(L \times L)$ 
blocks $\{\lambda\}$, the number
of single points (U)
and ``bonds'' (t or J)
that are being counted
is unique.
The ground state of the entire
system is trivially bounded
from above by 
\begin{eqnarray}
N \epsilon_{good},
\end{eqnarray}
with
\begin{eqnarray}
\langle \psi_{var}| h_{\lambda}| \psi_{var} \rangle   = \epsilon_{good},
\end{eqnarray}
where the expectation value is now evaluated for 
$| \psi_{var} \rangle$
on the open $L \times L$ cluster.

(Notwithstanding, when periodic boundary conditions
are employed, and the final periodic 
$L \times L$ system 
is unraveled to cover the 
entire lattice each site (with its
associated interaction
energy $U$ in the Hubbard
model), and each link
(with its associated kinetic 
energy (t) and (for the t-J
model) its associated
exchange energy J)
appear in exactly $L^{2}$
different blocks $\{\lambda\}$.)

Explicitly,
$\delta = L/(L+1) \times$  (the minimum on $\lambda$ 
with periodic boundary
conditions when 
also including bonds connecting
opposite boundaries) -
(open boundary condition minimum on $\lambda$ when
constrained to lie outside the ``good''
bar-code sector).

The factor of L/(L+1) originates from
the fact that when tiling the plane with clusters
$\lambda$ there are no interactions amongst opposite
sides of $\lambda$ that can be accounted for. 
The computation for the open and closed
boundary condition blocks are done on
equal footing.

If $\epsilon_{good}$ is of lower value
than the lower bounds (see the previous section)
which we will obtain for all
other ``bad'' sectors then
we immediately prove that
the ground state of the 
entire lattice is favorably
($N_{b} \le N_{g}$)
in the same sector 
as the periodic boundary
condition minimum.
That is, in each $L \times L$ 
window we will likely see the same 
pattern modulu trivial translations. 
If $\epsilon_{good}$ is not lower
than the lower bounds on the 
other sectors, we might need to
increase the size of our
$L \times L$ blocks
and re-compute upper and lower
bounds. 

That the two bounds should cross
is to be expected for coarse covering
as, in the large $L$ limit, 
the periodic boundary condition
state becomes the true ground
state of the system,
and changing it to a 
different sector 
(altering any given
plaquette(s) to a 
different bar code state)
will entail a finite
energy penalty.

\section{Putting all of the pieces together- 
Gaps and Antigaps}
\label{pieces}

If the energy gap separating 
the open boundary condition minimum
on all ``bad'' sectors from the
periodic boundary condition
minimum within the ``good'' 
sector is $\Delta$ and 
if the open boundary condition
minimum (within the ``good'' sector)
is lower by $\delta$ compared to
the periodic boundary condition
minimum, then if within the
maximal tiling of the plane
with all maximally overlapping
clusters, the number
of clusters lying within
the ``good'' sector is $N_{g}$
and the number of clusters 
lying within the ``bad'' sector
is $N_{b}$ then  the energy
difference between that 
configuration is elevated
by at least
\begin{eqnarray}
\overline{\Delta} \equiv [N_{b} \Delta - N_{g} \delta]
\end{eqnarray}
relative to the 
periodic boundary condition
minimum.

In the limit of large cluster
size, both the periodic boundary condition
minimum and the open 
boundary condition minimum 
must match with that 
of the global ground state energy 
state: $\delta \to 0$.

For large enough clusters, the lower bound, $\overline{\Delta}$,
on the energy penalty as compared
to the true global ground state 
becomes meaningful. Only bad clusters of number
\begin{eqnarray}
N_{b} \le (\frac{\delta}{\Delta}) N_{g} 
\label{center}
\end{eqnarray}
will not elevate the ground state 
according to our bound (i.e. will
not lead to $\overline{\Delta} >0$).
By their very definition, the gaps must satisfy
$\delta < \Delta$ if both the 
open and periodic boundary condition
minima give rise to the same bar coded state(s), 
and $N_{b}$ is always smaller than $N_{g}$.
The bounds become increasingly serious 
as the ration $(\delta/\Delta)$ diminishes.

Note that as the open
boundary condition
problem is not translationally
invariant, we will in general find 
patterns that are related to each
other by translations on the 
global
lattice yet have different
minimal energies on the
open boundary condition
problem on the fragment
(cluster). In such a 
case, the bounds will
even improve. If several
good patterns have 
relative ``anti-gaps''
$\delta_{i}$ then 
for those we will
replace $N_{g} \delta$
in the bound by the appropriate
sum of $\{\delta_{i}\}$. In
the most stringent bound, we 
may set, of course, $\delta 
\equiv \max_{i} \delta_{i}$. 

Even if we have a finite size block $\lambda$ for 
which the periodic boundary condition and 
the open boundary condition minima lie in
different abbreviated bar code sectors,
the more ``detailed  balance'' equation, 
\begin{eqnarray}
\sum_{\{\lambda_{i}\}} (\Delta_{i} - \delta_{i}) \le 0,
\label{severe}
\end{eqnarray}  
will lead to a strong constraint.

\section{A Possible Via For Proving Stripes On the Infinite Lattice}
\label{why}

\begin{figure}
\includegraphics[width=8cm]{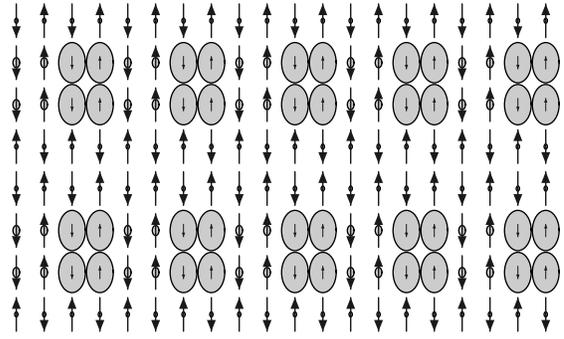}
\caption{A bond centered stripe as attained in mean 
field theory.
From M. Bosch et al. Earlier similar
pictures were found by 
the detailed
DMRG calculations of White and 
Scalapino.}   
\label{AAA4}
\end{figure}

If results similar to the end product 
of the DMRG calculations of White and Scalapino \cite{WS}
are found for a system with 
both open and closed boundary conditions
(similar to the pattern depicted
in Fig.(\ref{AAA4})) 
then the existence 
of stripes on the infinite lattice 
will effectively 
be proven for that Hamiltonian. All states
with densities in between (and including) 
the two extremes (the periodic and open boundary condition minima)
will occur (up to trivial translations)
at least 50\% of the time. The proof is 
trivial: both the open and 
closed boundary condition minima lie in the same sector
(this is how we readjusted the definition
of the ``good'' sector) and consequently $\delta  < \Delta$. 
Inserted in Eqn.(\ref{center}), this
demonstrates that $N_{b} \le N_{g}$-
i.e. ``good'' charge order will be 
observed in more than a 
half of the blocks $\{ \lambda \}$. 
In general, Eqn.(\ref{severe})
will impose restrictions
on the form of the ground state.

If the stripes found by DMRG
are a finite size artifact 
and indeed do result from the use 
of the open boundary conditions,
then we can add additional short Coulomb 
repulsions (always present
in the real system) to enhance 
non uniform charge order in both the open boundary condition
problem (which already displays non-uniform 
order without this enhancement) and within
the periodic boundary condition
problem and to establish rigorous
bounds on non-uniform charge and 
spin densities.

The real question is, of course, what 
system sizes are required before nonuniform
charge order might be observed within 
the ground state of a system subjected
to periodic boundary conditions. Periodic boundary conditions
merely favor certain commensurate Fourier 
modes: non-uniform configurations 
do, of course, arise. Nevertheless, generically,
charge density like oscillations 
within the ground state are far easier
to observe in systems
subject to open boundary
conditions. If the commensurability
is low, e.g. natural candidate
ground states for 1/8 doping, then 
one might naively expect the periodic ground 
state not to deviate by much relative 
to that on an open fragment.

{\bf{Acknowledgments}} 

It is a pleasure to thank Assa Auerbach 
for numerous conversations and 
correspondence.


\begin{thebibliography}{10}


\bibitem{WS} S.~ R.~ White and D.~ J.~ Scalapino,
Phys. Rev. Lett. {\bf 80}, 1272 (1998), cond-mat/9907375


\bibitem{stripes}  J.~ Zaanen and O.~ Gunnarsson, Phys. Rev. B {\bf 40}, 7391
(1989); K.~ Machida, Physica C {\bf 158}, 192 (1989); D.~ Poilblanc and
T.~ M.~ Rice, Phys. Rev. B {\bf 39}, 9749 (1989); H.~ J.~ Schulz,
J. Physique,
{\bf 50}, 2833 (1989); M.~ Kato, K.~ Machida,
H.~ Nakanishi, and M.~ Fujita, J. Phys. Soc. Jpn., {\bf 59}, 1047
(1990); J.~ A.~ Verges et al., Phys. Rev. B {\bf 43}, 6099 (1991);
M.~ Inui and P.~B.~ Littlewood, Phys. Rev. B {\bf 44}, 4415 (1991)


\bibitem{phase}
V.~ J.~ Emery and S.~ A.~ Kivelson, Physica C {\bf 209},
597 (1993)  


\bibitem{AAA} Ehud Altman and Assa Auerbach, 
Phys. Rev. B 65, 104508 (2002), cond-mat/0108087 


\bibitem{marco} M.~ Bosch, W.~ van~ Saarloos, and J.~ Zaanen,
Phys. Rev. B {\bf 63},92501 (2001),
cond-mat/0003236 



\end{thebibliography}
\end{document}